\documentclass[twocolumn,preprintnumbers,superscriptaddress,altaffilletter,nofootinbib,amsmath,amssymb]{revtex4}
\usepackage{graphicx}% Include figure files
\usepackage{dcolumn}% Align table columns on decimal point
\usepackage{bm}% bold math
\usepackage{xcolor}
\usepackage{epsfig,amsfonts,amstext,afterpage,psfrag}
\usepackage{slashed}

\usepackage[latin1]{inputenc}
\def\bea{\begin{eqnarray}} 
\def\eea{\end{eqnarray}}
\def\be{\begin{equation}} 
\def\ee{\end{equation}}

\newcommand{\beq}{\begin{equation}}
\newcommand{\eeq}{\end{equation}}

\newcommand{\tmfloatcontents}{}
\newlength{\tmfloatwidth}

\newcommand{\tmfloat}[5]{
	\renewcommand{\tmfloatcontents}{#4}
	\setlength{\tmfloatwidth}{\widthof{\tmfloatcontents}+1in}
	\ifthenelse{\equal{#2}{small}}
	{\ifthenelse{\lengthtest{\tmfloatwidth > \linewidth}}
		{\setlength{\tmfloatwidth}{\linewidth}}{}}
	{\setlength{\tmfloatwidth}{\linewidth}}  \begin{minipage}[#1]{\tmfloatwidth}
		\begin{center}
			\tmfloatcontents
			\captionof{#3}{#5}
		\end{center}
\end{minipage}}

\allowdisplaybreaks
\begin{document}
\preprint{SI-HEP-2021-24\\}
%%%%%%%%%%%%%%%%%%%%%%%%%%%%%%%%%

\title{Scalar resonances in the hadronic light-by-light contribution to the muon $(g-2)$} 
\author{Luigi Cappiello}
\affiliation{Dipartimento di Fisica \lq\lq Ettore Pancini'', Universit\'a di Napoli 'Federico II', Via Cintia, 80126 Napoli, Italy}
\affiliation{INFN-Sezione di Napoli, Complesso Universitario di Monte S. Angelo, Via Cintia Edificio 6, 80126 Napoli, Italy}

\author{Oscar Cat\`a}
\affiliation{Center for Particle Physics Siegen (CPPS), Theoretische Physik 1, Universit\"at Siegen, Walter-Flex-Stra\ss e 3, D-57068 Siegen, Germany}

\author{Giancarlo D'Ambrosio}
\affiliation{INFN-Sezione di Napoli, Complesso Universitario di Monte S. Angelo, Via Cintia Edificio 6, 80126 Napoli, Italy}

\begin{abstract}
\begin{center} {\bf Abstract}\end{center}
We evaluate the contribution of scalar mesons to the hadronic light-by-light piece of the muon anomalous magnetic moment, using a warped five-dimensional model and holographic methods. We assess the contribution of the lightest, sub-GeV scalars $\sigma(500)$, $a_0(980)$ and $f_0(980)$ together with their associated towers of excited states, which the model generates automatically. Our results point at a clearly negative contribution, overwhelmingly dominated by the $\sigma(500)$ meson, that we estimate at $a_\mu^{\rm{HLbL,S}}=-9(2)\cdot 10^{-11}$. This number is in very good agreement with the most recent determinations from dispersive analyses.  
\end{abstract}
\maketitle
\allowdisplaybreaks

%%%%%%%%%%%%%%%%%%%%%%%%%%%%%%%%%%%%%%%%%%%%%%%%%%%%%%%%%%%%%%%%%
\section{Introduction}
\label{sec:1}

Fermilab has recently released a measurement of the muon anomalous magnetic moment \cite{Abi:2021gix}, confirming the previous BNL E871 result \cite{Bennett:2006fi} and thus the disagreement with the theory expectation. Compared to the most recent theory determination \cite{Aoyama:2020ynm}, based on Refs. \cite{Aoyama:2012wk,Aoyama:2019ryr,Czarnecki:2002nt,Gnendiger:2013pva,Davier:2017zfy,Keshavarzi:2018mgv,Colangelo:2018mtw,Hoferichter:2019mqg,Davier:2019can,Keshavarzi:2019abf,Kurz:2014wya,Melnikov:2003xd,Masjuan:2017tvw,Colangelo:2017fiz,Hoferichter:2018kwz,Gerardin:2019vio,Bijnens:2019ghy,Colangelo:2019uex,Blum:2019ugy,Colangelo:2014qya}, the discrepancy is now pushed beyond four standard deviations. The Fermilab result is based on an analysis of the first data run, with a precision comparable to the one of the BNL number. In a matter of few years, the precision will be improved by a factor 4, thereby reaching the projected 0.14ppm.

On the theory side, the hadronic contributions are currently the focus of attention. The most pressing issue concerns the hadronic vacuum polarization contribution, where one needs to understand how compatible the increasingly precise lattice simulations (see, e.g., Ref. \cite{Borsanyi:2020mff}) are with the data-driven analyses \cite{Davier:2019can,Keshavarzi:2019abf} and low-energy phenomenology \cite{Crivellin:2020zul,Keshavarzi:2020bfy,Malaescu:2020zuc,Colangelo:2020lcg}. 

The hadronic light-by-light (HLbL) contribution is in a less critical stage, with a reasonable level of precision and with overall agreement between the different calculational techniques. Data-driven dispersive analysis~\cite{Colangelo:2014dfa,Colangelo:2014pva} and lattice simulations~\cite{Blum:2019ugy,Chao:2021tvp} still have room for improvement, and more precise determinations are expected to come in the near future. 

In parallel, there have been some recent developments for the HLbL using five-dimensional Lagrangian-based models with techniques borrowed from the AdS/CFT correspondence \cite{Maldacena:1997re,Gubser:1998bc,Witten:1998qj}. When compactified to four dimensions, these models can be interpreted as hadronic models of QCD with an infinite number of states for the different meson channels, as it is expected in the large-$N_c$ limit of QCD \cite{tHooft:1973alw,Witten:1979kh}. These models are provided with the following relevant features:
\begin{itemize}
\item[(i)] They undergo spontaneous chiral symmetry breaking, and thus they contain a pion multiplet, together with all the consequences of chiral dynamics built in; 
\item[(ii)] At high energies, they are conformal invariant and match the QCD asymptotic behaviour of correlators in the scaleless limit;
\item[(iii)] The chiral anomaly is consistently implemented at all energy scales. 
\end{itemize}

With these ingredients, these toy models are especially suited to address conceptual issues in simplified but consistent settings, especially issues related to the duality between hadronic contributions at high energies and QCD short-distance constraints. This was demonstrated recently in Refs. \cite{Leutgeb:2019gbz,Cappiello:2019hwh}, in which it was shown that a nontrivial combination of the Goldstone and axial-vector contributions, mostly dictated by anomaly matching at all energy scales, naturally resolved a long-standing puzzle \cite{Melnikov:2003xd,Colangelo:2019uex} regarding the saturation of short-distance constraints with hadronic states in the HLbL. Interestingly, the holographic models also predict a more sizeable axial-meson contribution than previously estimated (see, e.g., Ref. \cite{Colangelo:2021nkr} for a detailed comparison). This conclusion has been shown to be solid even in the presence of quark-mass corrections \cite{Leutgeb:2021mpu}.

So far, holographic models have been used to examine and evaluate Goldstone \cite{Hong:2009zw,Cappiello:2010uy,Leutgeb:2019zpq}, axial-vector \cite{Leutgeb:2019gbz,Cappiello:2019hwh,Leutgeb:2021mpu} and massive pseudoscalar \cite{Leutgeb:2021mpu} contributions. In this paper, we will apply the same formalism to evaluate the scalar contributions to the HLbL. There is generalized consensus that the scalar contributions are rather modest, to a large extent dominated by the $\sigma(500)$ resonance, and with the opposite sign as the Goldstone- and axial-vector contributions. Apart from the contribution of the lightest, sub-GeV scalars, our model allows us to estimate the contribution of the infinite towers of excited scalar states, and thereby get a rough idea of the uncertainties that can be ascribed to heavier, not experimentally accessible, scalars.  

By construction, our model treats the $\sigma(500)$ meson as a narrow-width state. Regardless of the nature of the $\sigma$ meson in the large-$N_c$ limit (see, e.g., Refs. \cite{RuizdeElvira:2010cs,Pelaez:2015qba} for a detailed discussion), at $N_c=3$ its width is simply too big to be described as a narrow-width resonance. This raises the question of whether its contribution to the HLbL, as predicted by the model, is reliable. In order to estimate the $\sigma(500)$ parameters (e.g., its mass and its decay width into photons), the most rigorous approach uses dispersive methods including coupled channels applied to $\gamma\gamma\to\pi\pi$ (see, e.g., Refs. \cite{Hoferichter:2011wk,Moussallam:2011zg,Hoferichter:2019nlq}) or, more recently, to $\gamma\gamma\to\pi\pi$ and $\gamma\gamma\to KK$ data \cite{Danilkin:2019opj,Danilkin:2020pak}. However, it would be inconsistent to use the values obtained with dispersive methods, which use a complex pole definition for the $\sigma$, in our approach, which is close to the Breit-Wigner definition. To the best of our knowledge, the set of values closest to our parametrization are the ones provided in Ref. \cite{Filkov:1998rwz} from a fit to $\gamma\gamma\to\pi\pi$ data. With these values, we estimate the $\sigma$ contribution to the HLbL at $a_\mu\sim-9\cdot 10^{-11}$, in excellent agreement with the value from the most recent dispersive analyses \cite{Colangelo:2017fiz,Danilkin:2021icn}.

The previous conceptual issue has a much milder effect for the $a_0(980)$ and $f_0(980)$, which can be considered narrow states to a good approximation. The values we find for those states can be jointly estimated as $a_\mu\sim-0.6\cdot 10^{-11}$, which are in agreement with Refs. \cite{Knecht:2018sci,Danilkin:2021icn}. We have also evaluated the contribution from the excited towers of scalar mesons, assuming them all as sufficiently narrow, and estimated their effect on the HLbL as $\Delta a_\mu^S\sim-0.2(2)\cdot 10^{-11}$. Summing up all the contributions, our final number for the total scalar contribution reads $a_\mu^S=-9(2)\cdot 10^{-11}$.

This paper is organized as follows: in Sec. \ref{sec:2} we lay out the details of our setup, which is then used in Sec. \ref{sec:3} to derive the expression of the hadronic four-point tensor entering the HLbL. In Sec. \ref{sec:4}, we derive the scalar transition form factor and the $\langle SVV\rangle$ correlator from the model and examine their behaviour at low and high energies. In Sec. \ref{sec:5}, we discuss our choice of parameters and compute the inclusive scalar contributions to the muon $g-2$. Conclusions and final remarks are given in Sec. \ref{sec:6}. 
    
%%%%%%%%%%%%%%%%%%%%%%%%%%%%%%%%%%%%%%%%%%%%%%%%%%%%%%%%%%%%%%%%%
\section{The model}
\label{sec:2}

In this section we will introduce the model we will be using and work out the fundamental ingredients that will be needed for the determination of the HLbL contribution. Sec. \ref{subsec:1} describes the general features of the model, while Secs. \ref{subsec:2} to \ref{subsec:4} are of a more technical nature and can be skipped on a first reading.
%%%%%%%%%%%%%%%%%%%%%%%%%%%%%%%%%%%%%%%%%%%%%%%%%%%%%%%%%%%%%%%%%
\subsection{From the five-dimensional action to a theory of hadrons}
\label{subsec:1}

The model we will use is a toy model of QCD in the large-$N_c$ limit, with infinite towers of vector, axial-vector, scalar and pseudoscalar mesons. Such a large-$N_c$ realization can be constructed in a very efficient and economical way from a five-dimensional formulation \cite{Sakai:2004cn,Erlich:2005qh,DaRold:2005zs,DaRold:2005vr,Hirn:2005nr}, by using the techniques developed in the context of the AdS/CFT correspondence \cite{Maldacena:1997re,Gubser:1998bc,Witten:1998qj}. The action we will use reads  
\begin{align}\label{action1}
S[L_M,R_M,X]&=\int d^5x\sqrt{g}\,{\cal{L}}(L_M,R_M,X)+S_{\rm{CS}}\,,
\end{align}
where
\begin{align}\label{action}
{\cal{L}}&(L_M,R_M,X)=-\lambda\,{\rm tr}\!\left[F_{(L)}^{MN}F_{(L)MN}+F_{(R)}^{MN}F_{(R)MN}\right]\nonumber\\
&+\rho\,{\rm tr}\!\left[D^{M}X^\dagger D_{M}X-m_X^2X^\dagger X \right]+z\,\delta(z-z_0)V(X)\nonumber\\
&+\zeta_+\,{\rm tr}\!\left[X^\dagger X F_{(R)}^{MN}F_{(R)MN}+X X^\dagger F_{(L)}^{MN}F_{(L)MN}\right]\nonumber\\
&+\zeta_-{\rm tr}\!\left[X^\dagger F_{(L)}^{MN}XF_{(R)MN}\right]
\end{align}
and 
\begin{align}
S_{\rm{CS}}=c\int\,{\rm tr}\big[\omega_5(L)-\omega_5(R) \big]\,.
\end{align}
The Chern-Simons term makes sure that the model has all the consequences of the axial anomaly built in. Its explicit expression in terms of the gauge fields reads $\omega_5(L)= L F_{(L)}^2+\tfrac{i}{2}
{L}^3F_{(L)}-\tfrac{1}{10} L^5$, where the wedge product of forms is implicitly understood. The same conventions and definitions apply to the right-handed sector.

We will take the five-dimensional background metric to be anti-de Sitter,
\begin{align}
g_{MN}dx^Mdx^N&=\frac{1}{z^2}\left(\eta_{\mu \nu}dx^{\mu}dx^{\nu} -
dz^2\right)\,,
\end{align} 
where $\mu,\nu=(0,1,2,3)$, $M,N=(0,1,2,3,z)$ and $\eta_{\mu\nu}$ has a mostly negative signature. For convenience, we have normalized the AdS curvature to unity. The fifth dimension is taken to be compact, with four-dimensional boundary branes at $(x^\mu,\epsilon)$ and $(x^\mu,z_0)$. 

The first line in Eq.~(\ref{action}) collects the Yang-Mills term for the gauge fields $L_M$ and $R_M$. Together with the Chern-Simons term, this was the action used in Ref.~\cite{Cappiello:2019hwh} to describe the contribution of axials and Goldstone bosons to the HLbL. In this work, we will adopt the same conventions, with $F_{(L)\,MN} =\partial_{M}L_{N}-\partial_{N}L_{M}-i[L_{M},L_{N}]$, $L_M=L_M^at^a$ and ${\mathrm{tr}}\,(t^a \,t^b)=\tfrac{1}{2}\delta_{ab}$, where $t^a$ are the eight Gell-Mann matrices extended with $t^0=\textbf{1}_3/\sqrt{6}$. 

The remaining lines in Eq.~(\ref{action}) collect the scalar sector, with the complex field $X=X^at^a$ transforming as a bifundamental of $U(3)_L\times U(3)_R$, $X\to g_L X g_R^{\dagger}$. Accordingly,
\begin{align}
D_MX=\partial_M X-iL_MX+iXR_M\,.
\end{align}
For reasons to be explained below, we have introduced a potential term, localized on the infrared boundary. As in Ref. \cite{DaRold:2005vr}, we will take it to be
\begin{align}
V(X)=\frac{1}{2}\mu^2\,{\rm tr}\!\left[X^\dagger X\right]-\eta\,{\rm tr}\!\left[(X^\dagger X)^2\right]\,.
\end{align}  
The operators in the last two lines of Eq.~(\ref{action}), of higher dimension, generate the $S\gamma\gamma$ interactions and are therefore essential to have a scalar contribution to the HLbL.

Note that the nonanomalous part of the Lagrangian is endowed with a parity symmetry ($L_M\leftrightarrow R_M, X\leftrightarrow X^\dagger$), which will be broken by the infrared boundary terms.

The previous action describes the interactions of vector, axial, scalar and pseudoscalar mesons, defined as $L_{M}=V_{M}-A_{M}$, $R_{M}=V_{M}+A_{M}$ and $X=S+iP$. Using the AdS/CFT correspondence~\cite{Maldacena:1997re}, one can construct the dual four-dimensional field theory on the $z=\epsilon$ boundary, where the fields at the boundary correspond to external sources~\cite{Gubser:1998bc, Witten:1998qj}. With the appropriate choice of boundary conditions for the gauge fields at $z=z_0$, to be discussed below, the model can be made to undergo spontaneous symmetry breaking with the pattern expected from large-$N_c$ QCD, namely $U(3)_L\times U(3)_R\to U(3)_V$. The spectrum of the four-dimensional theory contains infinite towers of vector, axial, scalar and pseudoscalar massive modes, which can be expressed in terms of $V_\mu$, $A_\mu$, $S$ and $P$. In this paper we will only need to consider the vector and scalar fields and, accordingly, we will restrict our discussion to these particular sectors. Their Kaluza-Klein decomposition is
\begin{align}\label{KKmode}
V_{\mu}(x,z)&=f_V^{(0)}(z)V_{\mu}^{(0)}(x)+\sum_{n=1}V_{\mu}^{(n)}(x)\varphi_n^V(z)\,,\nonumber\\
S(x,z)&=X_0(z)+f_S^{(0)}(z)S^{(0)}(x)+\sum_{n=1}S^{(n)}(x)\varphi_n^S(z)\,,
\end{align}
where $X_0(z)$ is the vacuum configuration, to be determined below. The zero modes correspond to non-normalizable solutions of the quadratic equations of motion and overlap with the external sources. Through boundary conditions one can make sure that they are nondynamical.

If one is interested in computing correlators, it is, however, more convenient to express the fields in an alternative way, making the four-dimensional sources explicit. In the following, we will work in position space for the compact dimension and in momentum space for the remaining dimensions. Upper-case letters will be used for spacelike momenta. One can then write
\begin{align}\label{decomposition}
&V_\mu(z,Q)=v(z,Q){\hat{v}}_\mu(Q);\nonumber\\
&S(z,Q)=X_0(z)+s(z,Q){\hat{s}}(Q)\,,
\end{align}
where it is understood that ${\hat{v}}_\mu(Q)$ is a transverse vector, i.e.,
\begin{align}
{\hat{v}}_\mu(Q)=\left[\eta_{\mu\nu}-\frac{q_{\mu}q_{\nu}}{q^2}\right] {\hat{v}}_\nu(Q)\,.
\end{align}
The hatted quantities ${\hat{v}}_\mu={\hat{v}}_\mu^a t^a$ and ${\hat{s}}={\hat{s}}^a t^a$ are identified with the classical sources associated to the chiral currents $j^a_{\mu}= \overline{q} \gamma_{\mu}t^a q$ and $j^a= \overline{q} t^a q$, respectively. 

The effective four-dimensional action can be determined once the functions $X_0(z)$, $v(z,Q)$ and $s(z,Q)$ are known. At leading order, they result from the solution of the classical equations of motion. Integrating out the fifth dimension then simply corresponds to plugging their on-shell values into the action and performing the corresponding integral over the fifth dimension. The end result is a four-dimensional generating functional, out of which the correlators of the theory can be computed. The $z=\epsilon$ and $z=z_0$ branes play the role of ultraviolet and infrared cutoffs of the four-dimensional theory, respectively.
%%%%%%%%%%%%%%%%%%%%%%%%%%%%%%%%%%%%%%%%%%%%%%%%%%%%%%%%%%%%%%%%%
\subsection{Vacuum configuration}
\label{subsec:2}

The vacuum of the theory can be found by solving the equations of motion at $p^2=0$, with gauge fields set to zero. This gives
\begin{align}
\left[\partial_z\left(\frac{1}{z^3}\partial_z \right)-\frac{m_X^2}{z^5}\right]X_0(z)=0\,,
\end{align}
whose solution is
\begin{align}
X_0(z)=A\left(\frac{z}{z_0}\right)^{\alpha_+}+B\left(\frac{z}{z_0}\right)^{\alpha_-}\,
\end{align}
with $\alpha_{\pm}=2\pm\sqrt{4+m_X^2}$. The masses of five-dimensional $p_j-$form fields are fixed by conformal invariance to satisfy
\begin{align}
m_j^2=(\Delta_j-p_j) (\Delta_j+p_j-d)\,,
\end{align}
where $\Delta_j$ is the conformal dimension of the currents associated to the different $p_j$-form fields. For QCD, these are the quark bilinears and therefore $\Delta_j=3$. Neglecting anomalous dimensions, this sets the gauge fields to be massless while $m_X^2=-3$. Within the AdS/CFT correspondence, the values of $X_0$ on the UV and IR branes are linked to explicit and spontaneous symmetry breaking, respectively~\cite{Klebanov:1999tb}. This suggests imposing \cite{Erlich:2005qh,DaRold:2005zs}
\begin{align}
X_0(\epsilon)=\epsilon {\cal{M}}_q\,;\qquad X_0(z_0)=s_0{\mathbf{1}}\,,
\end{align}
where ${\cal{M}}_q$ is the quark mass matrix and $s_0$ is a function of the parameters of the scalar sector, which can be determined by minimizing the action in the vacuum configuration. With these boundary conditions, the solution for the vacuum configuration reads
\begin{align}
X_0(z)=\frac{{\cal{M}}_qz_0^3-s_0\epsilon^2}{z_0(z_0^2-\epsilon^2)}z+\frac{s_0-{\cal{M}}_qz_0}{z_0(z_0^2-\epsilon^2)}z^3\,.
\end{align}
The parameter $s_0$ can be related to the quark condensate by matching onto the QCD vacuum. By inspecting the piece of the vacuum energy linear in ${\cal{M}}_q$ in both the model and QCD, one finds
\begin{align}\label{vacuum}
s_0=-\frac{\langle{\bar{q}}q\rangle z_0^3}{2\rho}\,.
\end{align}    
In the following, we will work in the chiral limit and set ${\cal{M}}_q=0$. The vacuum of the theory then takes the simple form
\begin{align}
X_0(z)=s_0\left(\frac{z}{z_0}\right)^3\,,
\end{align}
where we have taken the limit $\epsilon\to 0$. Minimizing the action in these limits yields
\begin{align}
s_0^2=\frac{\mu^2-6\rho}{4\eta}\,.
\end{align} 
%%%%%%%%%%%%%%%%%%%%%%%%%%%%%%%%%%%%%%%%%%%%%%%%%%%%%%%%%%%%%%%%%
\subsection{Bulk-to-boundary propagators}
\label{subsec:3}

We now turn our attention to the functions $v(x,z)$ and $s(x,z)$, which are determined by solving the five-dimensional equations of motion quadratic in the vector and scalar fields, respectively. They lead to the differential equations:
\begin{align}
\left[\partial_z\left(\frac{1}{z}\partial_z\right)-\frac{Q^2}{z}\right]v(z,Q)=0\,,\nonumber\\
\left[\partial_z\left(\frac{1}{z^3}\partial_z\right)-\frac{Q^2}{z^3}-\frac{m_X^2}{z^5}\right]s(z,Q)=0\,.
\end{align}
The boundary conditions at $z=\epsilon$ define the external sources. Using Eq.~(\ref{decomposition}), we will accordingly choose 
\begin{align}
v(\epsilon,Q)=1;\qquad s(\epsilon,Q)=\epsilon\,.
\end{align}
The boundary conditions at $z=z_0$ have to guarantee the absence of unwanted boundary terms of infrared origin. For the vector field, they have to be compatible with the pattern of symmetry breaking $U(3)_L\times U(3)_R\to U(3)_V$. Both requirements are achieved with 
\begin{align}
\partial_z v(z,Q)\bigg|_{z_0}=0;\qquad \left(\rho \partial_z+\frac{\nu}{z}\right)s(z,Q)\bigg|_{z_0}=0\,,
\end{align}
where $\nu=4\eta s_0^2-3\rho$. The solutions are linear combinations of modified Bessel functions,
\begin{align}
v(z,Q)&=Qz\left[K_1(Qz)+\frac{K_0(Q z_0)}{I_0(Q z_0)} I_1(Qz)\right]\,;\label{btobpropV}\\
s(z,Q)&=Qz^2\left[K_1(Qz)-\xi\frac{K_1(Q z_0)}{I_1(Q z_0)} I_1(Qz)\right]\,,\label{btobpropS}
\end{align}
where
\begin{align}
\xi&=\frac{\gamma-\alpha_K}{\gamma+\alpha_I};\quad &\alpha_K&=Qz_0\frac{K_0(Q z_0)}{K_1(Q z_0)};\nonumber\\
\gamma&=\frac{\rho+\nu}{\rho};\quad &\alpha_I&=Qz_0\frac{I_0(Q z_0)}{I_1(Q z_0)}\,.
\end{align} 
Note that the limit $\eta\gg 0$, $\mu\gg 0$ with $\mu^2/\eta=$ ct., which corresponds to $\nu\gg 0$, simplifies the scalar infrared boundary condition to $s(z_0,Q)=0$. The corresponding solution for the scalar function is found by taking the limit $\xi=1$. This limiting case, which reduces the potential to a constant, was used in Ref. \cite{Domenech:2010aq}. Here, we will consider the generic case. In the next subsection, we will show that the parameter $\gamma$ controls the position of the scalar meson masses. This feature will be used in Sec. \ref{sec:5} to fit the lightest scalar masses. 

%%%%%%%%%%%%%%%%%%%%%%%%%%%%%%%%%%%%%%%%%%%%%%%%%%%%%%%%%%%%%%%%%  
\subsection{Scalar wavefunctions and Green's function}
\label{subsec:4}

Next, we examine the normalizable solutions of the scalar quadratic equations of motion, which describe the meson interactions and are needed to evaluate the scalar meson-exchange contribution to the HLbL. The scalar wavefunctions are given by the equation
\begin{align}
\left[\partial_z\left(\frac{1}{z^3}\partial_z\right)-\frac{m_X^2}{z^5}\right]\varphi_n^S(z)=-\frac{m_n^2}{z^3}\varphi_n^S(z)\,,
\end{align}
subject to the boundary conditions:
\begin{align}\label{boundS}
\varphi_n^S(\epsilon)=0;\qquad \left(\rho \partial_z+\frac{\nu}{z}\right)\varphi_n^S(z)\bigg|_{z_0}=0\,.
\end{align}
The solution is                           
\begin{align}
\varphi^S_n(z)=A_n z^2J_1(m_n z)\,,
\end{align}
where $A_n$ is a constant which, from the orthonormalization condition 
\begin{align}
\rho \int_{\epsilon}^{z_0}\frac{dz}{z^3}\varphi_n^S(z)\varphi_m^S(z)=\delta_{nm}\,,
\end{align}
can be expressed as ($\omega_n\equiv m_n z_0$)
\begin{align}
A_n^2=\frac{2\omega_n}{\rho z_0^2}\left[\omega_n(J_0^2(\omega_n)+J_1^2(\omega_n))-2J_0(\omega_n)J_1(\omega_n)\right]^{-1}\,.
\end{align}
The boundary conditions in Eq. (\ref{boundS}) determine the position of the scalar masses, which are implicitly given by the condition
\begin{align}\label{spectrumS}
\gamma J_1(\omega_n)+\omega_nJ_0(\omega_n)=0\,.
\end{align}
With it, the normalization constant can be simplified to
\begin{align}\label{An}
A_n^2=\frac{2}{\rho z_0^2J_0^2(\omega_n)}\frac{\gamma^2}{\gamma^2+2\gamma+\omega_n^2}\,.
\end{align}

Another quantity of interest is the scalar Green's function, which is the solution of the differential equation
\begin{align}
\left[\partial_z\left(\frac{1}{z^3}\partial_z\right)-\frac{Q^2}{z^3}-\frac{m_X^2}{z^5}\right]G(z,z',Q)=-\frac{\delta(z-z')}{\rho}\,,
\end{align}
with the same boundary conditions as the wavefunctions. Taking $\epsilon\to 0$, the solution is
\begin{align}\label{GreenS}
G(z,z';Q)&=\frac{z^2z'^2}{\rho}I_1(Qz')\left[K_1(Qz)-\xi\frac{K_1(Q z_0)}{I_1(Q z_0)} I_1(Qz)\right]\nonumber\\
&\times\theta(z-z')+(z\leftrightarrow z')\,.
\end{align}
Alternatively, the Green's function can be expressed in terms of the wavefunctions, in the form of a spectral decomposition:
\begin{align}\label{Greenspectral}
G(z,z';q)=\sum_n^{\infty}\frac{\varphi_n^S(z)\varphi_n^S(z')}{m_n^2-q^2}\,.
\end{align}
Eq.~(\ref{GreenS}) contains the regularized sum of the infinite tower of scalar mesons and it is the most convenient form to determine the inclusive scalar contribution to the HLbL. The spectral representation is particularly useful to determine the contribution of single mesons. We will use both forms in Sec. \ref{sec:5}.
  
It only remains to determine the parameters of the model. Since we are interested in a determination of the HLbL, it is convenient to fix the parameters by matching to the most relevant quantities. In the next sections, we will work out the expression for the scalar contribution to the HLbL, the scalar transition form factors and the $\langle SVV\rangle$ correlator. The criteria to fix the parameters will be discussed in Sec.~\ref{sec:5}. 

%%%%%%%%%%%%%%%%%%%%%%%%%%%%%%%%%%%%%%%%%%%%%%%%%%%%%%%%%%%%%%%%%
\section{The hadronic light-by-light tensor}
\label{sec:3}

The fundamental object for the HLbL is the electromagnetic four-point correlator
\begin{align}
&\Pi^{\mu\nu\lambda\rho}(q_1,q_2,q_3) = -i \int d^4x\ d^4y\ d^4z \ e^{-i(q_1 \cdot x + q_2 \cdot y + q_3 \cdot z)}\nonumber\\
&\times \langle 0 | T \{ j_\mathrm{em}^\mu(x) j_\mathrm{em}^\nu(y) j_\mathrm{em}^\lambda(z) j_\mathrm{em}^\rho(0) \} | 0 \rangle\,, \label{HLbLTensorDefinition}
\end{align}
where $j_\mathrm{em}^\mu(x)={\bar{q}}\gamma^{\mu}{\hat{Q}}q$, with ${\hat{Q}}=\tfrac{1}{3}{\mathrm{diag}}(2,-1,-1)$ being the electromagnetic charge matrix. Our conventions for momenta are such that $q_1+q_2+q_3+q_4=0$.

The hadronic light-by-light tensor satisfies the Ward identities,
\begin{align}
\left\{q_1^{\mu},q_2^{\nu},q_3^{\lambda},q_4^{\rho}\right\}\times\Pi_{\mu\nu\lambda\sigma}(q_1,q_2,q_3)=0\,,
\end{align}
which bring the number of 138 independent kinematic invariants down to 43 gauge-invariant tensor structures \cite{Leo:1975fb,Colangelo:2015ama}. 

\begin{figure}[t]
\begin{center}
\includegraphics[width=0.47\textwidth]{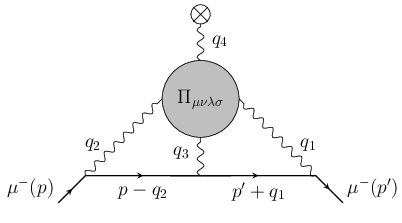}
\end{center}
\caption{The HLbL diagram. The blob represents the HLbL tensor. In our conventions, photon momenta are pointing inwards.} \label{fig:1}
\end{figure}

\begin{figure}[t]
\begin{center}
\includegraphics[width=0.5\textwidth]{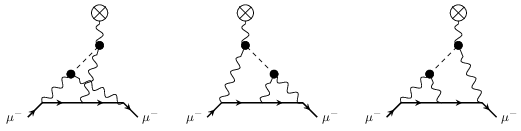}
\end{center}
\caption{The three scalar-exchange diagrams contributing to the HLbL. The dashed line denotes the scalar propagator and the black dots the scalar transition form factors. Photon momenta assignments are the same as in Fig.\ref{fig:1}, i.e., they point inwards.} \label{fig:2}
\end{figure}

Using projection techniques, the two-loop diagram of Fig.~\ref{fig:1} can be related to the muon anomalous magnetic moment as follows \cite{Aldins:1970id}:
\begin{align}\label{amu}
a_\mu^{\mathrm{HLbL}}&=-\frac{e^6}{48 m_\mu}\int \frac{d^4q_1}{(2\pi)^4} \frac{d^4q_2}{(2\pi)^4} {\cal{K}}_{\mu\nu\lambda\rho}\nonumber\\
&\times \left( \frac{\partial}{\partial q_4^\rho} \Pi^{\mu\nu\lambda\rho}(q_1,q_2,-q_4-q_1-q_2) \right) \bigg|_{q_4=0}\,,
\end{align}
with
\begin{widetext}
\begin{align}
{\cal{K}}_{\mu\nu\lambda\rho}=\frac{{\mathrm{tr}}\left( (\slashed p + m_\mu) [\gamma^\rho,\gamma^\sigma] (\slashed p + m_\mu) \gamma^\mu (\slashed p + \slashed q_1 + m_\mu) \gamma^\lambda (\slashed p - \slashed q_2 + m_\mu) \gamma^\nu \right)}{q_1^2 q_2^2 (q_1+q_2)^2[(p+q_1)^2 - m_\mu^2][(p-q_2)^2 - m_\mu^2]}\,.
\end{align}
\end{widetext}
By decomposing the HLbL tensor, one can write Eq. (\ref{amu}) in a simplified form. In this paper, we will use the formalism developed in Refs. \cite{Colangelo:2015ama,Colangelo:2017fiz} for dispersion analyses, which has become standard. Building on the strategy outlined in Refs. \cite{Bardeen:1969aw,Tarrach:1975tu}, the hadronic light-by-light tensor can be decomposed in gauge-invariant tensor structures free from singularities and zeros, at the expense of enlarging the basis from 43 to 54 elements:
\begin{align}
\Pi^{\mu\nu\lambda\rho}=\sum_{i=1}^{54} {\hat{T}}_i^{\mu\nu\lambda\rho}{\hat{\Pi}}_i\,.
\end{align}                                             
Out of these tensors, only 19 structures contribute to $a_\mu$. For the scalar exchange diagrams depicted in Fig. \ref{fig:2}, one easily concludes that only six of the structures are in general nonzero. 

In our model, the computation of the diagrams in Fig. \ref{fig:2} requires solving the scalar equation of motion. From the action of Eq. (\ref{action1}), this reads
\begin{align}
&\left[\partial_z\left(\frac{1}{z^3}\partial_z\right)-\frac{Q^2}{z^3}-\frac{m_X^2}{z^5}\right]s^a(z,Q)=-\frac{4X_0(z')}{z'}\zeta\nonumber\\
&\qquad\qquad\qquad\times{\mathrm{tr}}\bigg[t^a(F_{\mu\nu}F^{\mu\nu}-2\partial_{z'} V_{\rho}\partial_{z'} V^{\rho})\bigg]\,.
\end{align}
where $\zeta=\zeta_++\tfrac{1}{2}\zeta_-$. The solution can be found perturbatively as $s^a(z,Q)=s_{(0)}^a(z,x)+s_{(1)}^a(z,x)+\dots$, assuming, as usual, that interaction terms are small corrections to the kinetic term. The first-order function $s_{(0)}^a(z,x)$ solves the homogeneous equation and is nothing else than the bulk-to-boundary propagator of Eq. (\ref{btobpropS}). The second-order solution reads
\begin{align}
s_{(1)}^a(z,x)&=\int_{\epsilon}^{z_0} dz'G^{(a)}(z,z';x)\nonumber\\
&\times\left\{\frac{4X_0(z')}{z'}\zeta{\mathrm{tr}}\bigg[t^a(F_{\mu\nu}F^{\mu\nu}-2\partial_{z'} V_{\rho}\partial_{z'} V^{\rho})\bigg]\right\}\,,
\end{align}
and is the one relevant to generate the diagrams in Fig. \ref{fig:2}. Inserting the previous solution into the action in Eq. (\ref{action1}), the relevant operators read
\begin{align}\label{VVVV}
{\cal{L}}_{\rm{eff}}^{\rm{HLbL(scalar)}}&=\int_{\epsilon}^{z_0} dz {\cal{R}}^{(a)}(z,x)\nonumber\\
&\times\int_{\epsilon}^{z_0} dz'  G^{(a)}(z,z^{\prime};x) {\cal{R}}^{(a)}(z^{\prime},x)\,, 
\end{align}
where summation over the flavour index $a=0,3,8$ is understood and
\begin{align}
{\cal{R}}^{(a)}(z,x)=2\sqrt{2}\zeta{\hat{d}}^{a\gamma\gamma}\frac{X_0(z)}{z}\bigg[\frac{1}{2}F_{\mu\nu}F^{\mu\nu}-\partial_z V_{\rho}\partial_z V^{\rho}\bigg]\,.
\end{align}
The notation ${\hat{d}}^{abc}={\mathrm{tr}}[t^a\{t^b,t^c\}]$ and ${\hat{d}}^{a\gamma\gamma}=2{\mathrm{tr}}[t^a{\hat{Q}}^2]$ will be used henceforth.

The HLbL tensor then takes the form
\begin{widetext}
\begin{align}\label{VVVVM}
\Pi_{\mu\nu\lambda\rho}&(q_1,q_2,q_3,q_4)=\int_{\epsilon}^{z_0} dz \int_{\epsilon}^{z_0} dz'\left[T^{\mu\nu(a)}_{12}G^{(a)}(z,z^{\prime};s)T^{\lambda\rho(a)}_{34}+T^{\mu\lambda(a)}_{13}G^{(a)}(z,z^{\prime};t)T^{\nu\rho(a)}_{24}+T^{\mu\rho(a)}_{14}G^{(a)}(z,z^{\prime};u)T^{\nu\lambda(a)}_{23}\right]\,,
\end{align}
\end{widetext}
where we have used the Mandelstam variables $s=(q_1+q_2)^2$, $t=(q_1+q_3)^2$, $u=(q_1+q_4)^2$. The dynamical tensors $T^{\mu\nu(a)}_{ij}$ are defined as 
\begin{align}
T^{\mu\nu(a)}_{ij}(z)={\cal{P}}^{(a)}_{ij}(z)P^{\mu\nu}(q_i,q_j)+{\cal{Q}}^{(a)}_{ij}(z)Q^{\mu\nu}(q_i,q_j)\,,
\end{align}
with the gauge-invariant tensors given by
\begin{align}\label{gaugeinvT}
P^{\mu\nu}(q_1,q_2)&=q_2^{\mu}q_1^{\nu}-(q_1\cdot q_2)\eta^{\mu\nu}\,,\nonumber\\
Q^{\mu\nu}(q_1,q_2)&=q_2^2q_1^{\mu}q_1^{\nu}+q_1^2q_2^{\mu}q_2^{\nu}-(q_1\cdot q_2)q_1^{\mu}q_2^{\nu}-q_2^2q_1^2\eta^{\mu\nu}\,,
\end{align}
and
\begin{align}
{\cal{P}}^{(a)}_{ij}(z)&=8\zeta{\hat{d}}^{a\gamma\gamma}\frac{X_0(z)}{z}v(z,q_i)v(z,q_j)\,,\nonumber\\
{\cal{Q}}^{(a)}_{ij}(z)&=8\zeta{\hat{d}}^{a\gamma\gamma}\frac{X_0(z)}{z}\frac{\partial_z v(z,q_i)}{q_i^2}\frac{\partial_z v(z,q_j)}{q_j^2}\,.
\end{align}
Notice that in our model the two form factors above stem from the same combination of five-dimensional operators. Their expressions are a consequence of Lorentz invariance in five dimensions and our choice of AdS metric. However, there is no reason why the form factors should possess this underlying structure. Actually, we will show later on that these constraints inherited from the structure of the model are not a property of QCD.
   
The decomposition of the contribution to the HLbL in the 19-structure set gives
\begin{align}\label{scalarF}
{\hat{\Pi}}_4&=\int_{\epsilon}^{z_0}dz\left[{\cal{P}}^{(a)}_{12}(z)+(Q_1^2+Q_2^2+Q_1\cdot Q_2){\cal{Q}}^{(a)}_{12}(z)\right]\nonumber\\
&\times \int_{\epsilon}^{z_0}dz'G_{(a)}(z,z';s){\cal{P}}^{(a)}_{34}(z')\,,\nonumber\\
{\hat{\Pi}}_{17}&=\int_{\epsilon}^{z_0}dz{\cal{Q}}^{(a)}_{12}(z)\int_{\epsilon}^{z_0}dz'G_{(a)}(z,z';s){\cal{P}}^{(a)}_{34}(z')\,,
\end{align}
where the scalar invariants are numbered according to the notation introduced in Ref. \cite{Colangelo:2017fiz}. The remaining four invariants, namely ${\hat{\Pi}}_5$, ${\hat{\Pi}}_6$ and ${\hat{\Pi}}_{11}$, ${\hat{\Pi}}_{16}$ in the same notation, can be found by symmetric permutation of ${\hat{\Pi}}_4$ and ${\hat{\Pi}}_7$, respectively. We note, as observed before \cite{Knecht:2018sci}, that the contributions proportional to the form factor products ${\cal{Q}}^{(a)}_{ij}(z){\cal{Q}}^{(a)}_{kl}(z')$ do not contribute to $a_\mu$.

Five of the integrals in Eq. ({\ref{amu}}) can be performed independently of the form of the scalar functions. The remaining integrals can be written as a master formula for the HLbL, 
\begin{align}\label{master1}
	a_\mu^\mathrm{HLbL}&=\frac{2 \alpha^3}{3 \pi^2} \int_0^\infty dQ_1 \int_0^\infty dQ_2 \int_{-1}^1 d\tau \nonumber\\
&\times \sqrt{1-\tau^2} Q_1^3 Q_2^3\sum_{i=1}^{12} \bar T_i(Q_1,Q_2,\tau) \bar \Pi_i(Q_1,Q_2,\tau)\,,
\end{align}
where $Q_j$ is a shorthand notation for the moduli of Euclidean momenta and $\tau$ is the cosine of the angle between $Q_1^\mu$ and $Q_2^\mu$. The definitions of the generic kernels $\bar T_i(Q_1,Q_2,\tau)$ can be found in Ref. \cite{Colangelo:2017fiz}. For the scalar dynamical functions, we find
\begin{align}\label{12Basis}
{\bar\Pi}_3(Q_1,Q_2,\tau)&=\int_{\epsilon}^{z_0}dz\left[{\cal{P}}^{(a)}_{12}+(Q_1^2+Q_2^2+Q_1Q_2\tau){\cal{Q}}^{(a)}_{12}\right]\nonumber\\
&\times \int_{\epsilon}^{z_0}dz'G_{(a)}(z,z';s){\cal{P}}^{(a)}_{34}\,,\nonumber\\
{\bar\Pi}_4(Q_1,Q_2,\tau)&=\int_{\epsilon}^{z_0}dz\left[{\cal{P}}^{(a)}_{13}+(Q_1^2+Q_2^2+Q_1Q_2\tau){\cal{Q}}^{(a)}_{13}\right]\nonumber\\
&\times \int_{\epsilon}^{z_0}dz'G_{(a)}(z,z';t){\cal{P}}^{(a)}_{24}\,,\nonumber\\
{\bar\Pi}_8(Q_1,Q_2,\tau)&=\int_{\epsilon}^{z_0}dz{\cal{P}}^{(a)}_{14}\int_{\epsilon}^{z_0}dz'G_{(a)}(z,z';u){\cal{Q}}^{(a)}_{23}\,,\nonumber\\
{\bar\Pi}_9(Q_1,Q_2,\tau)&=\int_{\epsilon}^{z_0}dz{\cal{Q}}^{(a)}_{12}\int_{\epsilon}^{z_0}dz'G_{(a)}(z,z';s){\cal{P}}^{(a)}_{34}\,.
\end{align} 

The previous expressions are functions of the quantities determined in Sec.~\ref{sec:2}, which in turn are functions of the parameters of the model. Once the parameters are fixed, the scalar contribution to the HLbL can be straightforwardly computed. The parameters can be determined by matching onto some of the properties of the electromagnetic scalar transition form factor and the three-point correlator $\langle SVV\rangle$. 

%%%%%%%%%%%%%%%%%%%%%%%%%%%%%%%%%%%%%%%%%%%%%%%%%%%%%%%%%%%%%%%%%
\section{Scalar transition form factors and $\langle SVV\rangle$ correlator}
\label{sec:4}

The scalar transition from factors for the different scalar meson with flavour index $a$ are defined as
\begin{align}
&\Gamma_{\mu\nu}^{(n,a)}(q_1,q_2)=i\int d^4x\ e^{-iq_1 \cdot x} \langle 0 | T \{ j_\mathrm{em}^\mu(x) j_\mathrm{em}^\nu(0) \} | S_n^a \rangle\nonumber\\
&\quad=F_1^{(n,a)}(q_1^2,q_2^2)P_{\mu\nu}(q_1,q_2)+F_2^{(n,a)}(q_1^2,q_2^2)Q_{\mu\nu}(q_1,q_2)\,, \label{VVSDefinition}
\end{align}
where $P_{\mu\nu}$ and $Q_{\mu\nu}$ are the gauge-invariant tensors defined in Eq. (\ref{gaugeinvT}) and the superscript $n$ labels the excitation mode of the scalar tower. 

In our model, the form factors follow from the operators in the last two lines of Eq.~(\ref{action}), by decomposing the vector fields in terms of electromagnetic field sources, and expanding the scalar field in terms of its Kaluza-Klein modes, as in Eq.~(\ref{KKmode}). This yields transition form factors for each scalar meson, with the following expressions for the dynamical Lorentz-invariant structures:
\begin{align}\label{formfactors}
F_1^{(n,a)}(q_1^2,q_2^2)&=8\zeta{\hat{d}}^{a\gamma\gamma}\int_{\epsilon}^{z_0}dz\frac{X_0(z)}{z}\varphi_n^S(z)v_1(z)v_2(z)\,,\nonumber\\
F_2^{(n,a)}(q_1^2,q_2^2)&=8\zeta{\hat{d}}^{a\gamma\gamma}\int_{\epsilon}^{z_0}dz\frac{X_0(z)}{z}\varphi_n^S(z)\frac{\partial_z v_1(z)}{q_1^2}\frac{\partial_z v_2(z)}{q_2^2}\,.
\end{align} 

The scalar transition form factor predicted by the model can be compared, for on-shell photons, to the information that exists at low energies on the partial decay widths of scalars into two photons which, at least for the lightest scalars, can be determined from experiment. At high photon virtualities, the form factor can be instead determined using the OPE. 

Since $F_2^{(n,a)}(q_1^2,q_2^2)$ only contributes to processes with virtual photons, the decay width of the scalar into two on-shell photons can be expressed in terms of $F_1^{(n,a)}(0,0)$ alone as
\begin{align}\label{widths}
\Gamma_{\gamma\gamma}^{(n,a)}=\frac{\pi\alpha^2}{4}m_n^3\left|F_1^{(n,a)}(0,0)\right|^2\,.
\end{align}
From Eq.~(\ref{formfactors}), and using that $v(z,0)=1$, one finds
\begin{align}\label{F10}
F_1^{(n,a)}(0,0)&=8\zeta{\hat{d}}^{a\gamma\gamma}\int_{\epsilon}^{z_0}dz\frac{X_0(z)}{z}\varphi_n^S(z)\nonumber\\
&=8s_0 z_0^2\zeta{\hat{d}}^{a\gamma\gamma}\frac{A_n}{\omega_n^5}\int_0^{\omega_n}dy y^4 J_1(y)\nonumber\\
&=8s_0 z_0^2\zeta{\hat{d}}^{a\gamma\gamma}\frac{A_n}{\omega_n^2}\left[4J_3(\omega_n)-\omega_n J_4(\omega_n)\right]\,.
\end{align}
The limit of highly virtual photons can be computed by noting that for large $Q$, $v(z,Q)\sim Qz K_1(Qz)$. It is customary to define the variables $Q^2=\tfrac{1}{2}(Q_1^2+Q_2^2)$ and $w=(Q_1^2-Q_2^2)(Q_1^2+Q_2^2)^{-1}$, such that $Q_{1,2}=Q\sqrt{1\pm w}$. The model then predicts
\begin{align}\label{asymptoticsF1F2}
\lim_{Q^2\to\infty} F_1^{(n,a)}(Q_1^2,Q_2^2)&=\frac{1536}{35}\zeta{\hat{d}}^{a\gamma\gamma}\frac{s_0}{z_0^4}\frac{A_n\omega_n}{Q^6}f_1(w)\,,\nonumber\\
\lim_{Q^2\to\infty} F_2^{(n,a)}(Q_1^2,Q_2^2)&=\frac{1152}{35}\zeta{\hat{d}}^{a\gamma\gamma}\frac{s_0}{z_0^4}\frac{A_n\omega_n}{Q^8}f_2(w)\,,
\end{align}
with 
\begin{align}
&f_1(w)=\frac{35}{384}\sqrt{1-w^2}\int_0^\infty dy y^7 K_1(y\sqrt{1+w})K_1(y\sqrt{1-w})\nonumber\\
&\quad=\frac{35}{32w^7}\left[30w-26w^3-3(w^4-6w^2+5)\log\left(\frac{1+w}{1-w}\right)\right]\,,\nonumber\\
&f_2(w)=\frac{35}{288}\int_0^\infty dy y^7 K_0(y\sqrt{1+w})K_0(y\sqrt{1-w})\nonumber\\
&\quad=\frac{35}{12w^7}\left[-15w+4w^3-\frac{9w^2-15}{2}\log\left(\frac{1+w}{1-w}\right)\right]\,.
\end{align}
These expressions are regular in the symmetric limit, $w=0$, and normalized such that $f_1(0)=f_2(0)=1$. 

We note that the asymptotic behaviour of the scalar transition form factor predicted by the model, $\Gamma_{\mu\nu}^{(n,a)}(q_1,q_2)\sim Q^{-6}\Gamma_{\mu\nu}^{(\infty)}$, with $\Gamma_{\mu\nu}^{(\infty)}$ a tensorial structure quadratic in the hard momenta, does not match the one expected from short-distance QCD, where asymptotically one finds $\Gamma_{\mu\nu}^{(n,a)}(q_1,q_2)\sim Q^{-4}\Gamma_{\mu\nu}^{(\infty)}$~\cite{Knecht:2018sci}. If the comparison is made at the form factor level, where short-distance QCD predicts the asymptotic scalings $F_1(Q^2,Q^2)\sim Q^{-2}$ and $F_2(Q^2,Q^2)\sim Q^{-4}$~\cite{Knecht:2018sci,Hoferichter:2020lap}, the discrepancy is even more pronounced. However, the corresponding impact on the HLbL should be rather modest. In the next section we will provide some arguments in this direction. More important is the fact that Eqs.~(\ref{asymptoticsF1F2}) predict the same sign for both form factors at large Euclidean momenta. Both these limitations could in principle be amended with a more sophisticated action, e.g., the asymptotic sign between the form factors can be fixed with the addition of higher-derivative operators in the bulk such as ${\rm tr}\!\left[XX^\dagger \nabla^AF_{(L)AN}\nabla_BF_{(L)}^{BN}\right]$, while the high-energy behaviour can be improved with boundary operators. In this paper, we will not pursue these extensions and we will stick to the minimal action. 

A similar large-$Q^2$ falloff is found in the model when one of the photons is on shell and one highly off shell. In this case,
\begin{align}
\lim_{Q^2\to\infty} F_1^{(n,a)}(Q^2,0)&=1536\zeta{\hat{d}}^{a\gamma\gamma}\frac{s_0}{z_0^4}\frac{A_n\omega_n}{Q^6}\,,\nonumber\\
\lim_{Q^2\to\infty} F_2^{(n,a)}(Q^2,0)&=9216\zeta{\hat{d}}^{a\gamma\gamma}\frac{s_0}{z_0^4}A_n\omega_n\frac{\log (Qz_0)}{Q^8}\,.
\end{align}

We will now turn our attention to the $\langle SVV\rangle$ correlator, defined as
\begin{align}
&\Gamma_{\mu\nu}^{(a)}(q_2,q_2)=i^2\!\!\int d^4x\int d^4y \ e^{-i(q_1 \cdot x + q_2 \cdot y)}\nonumber\\
&\qquad\times\langle 0 | T \{ j_\mathrm{em}^\mu(x) j_\mathrm{em}^\nu(y) j_S^a(0) \} | 0 \rangle\nonumber\\
&\quad={\bar{{\cal{P}}}}^{(a)}(q_1^2,q_2^2)P_{\mu\nu}(q_1,q_2)+{\bar{{\cal{Q}}}}^{(a)}(q_1^2,q_2^2)Q_{\mu\nu}(q_1,q_2)\,. \label{VVSCDefinition}
\end{align}
In the model, the scalar invariants take the form
\begin{align}
{\bar{{\cal{P}}}}^{(a)}(q_1^2,q_2^2)&=8\zeta{\hat{d}}^{a\gamma\gamma}\int_{\epsilon}^{z_0}\frac{dz}{z}X_0(z)s_3(z)v_1(z)v_2(z)\,,\nonumber\\
{\bar{{\cal{Q}}}}^{(a)}(q_1^2,q_2^2)&=8\zeta{\hat{d}}^{a\gamma\gamma}\int_{\epsilon}^{z_0}\frac{dz}{z}X_0(z)s_3(z)\frac{\partial_z v_1(z)}{q_1^2}\frac{\partial_z v_2(z)}{q_2^2}\,.
\end{align}
We will first consider the limit when all momenta are much larger than the confinement scale. To leading order, one can consider $q_1=q_2=\tfrac{1}{2}q_3\equiv q$. The expressions above then yield
\begin{align}\label{SDCmodel}
\lim_{q^2\to\infty}\Gamma^{(a)}_{\mu\nu}(q,q)&=\frac{16s_0\zeta}{z_0^3}\frac{{\hat{d}}^{a\gamma\gamma}}{Q^4}(q_\mu q_\nu-q^2\eta_{\mu\nu})\nonumber\\
&\times\int_0^{\infty}dy y^6 K_1(2y)\left[K_1^2(y)-K_0^2(y)\right]\,.
\end{align}
This is to be compared with the QCD result, which follows from an application of the OPE. The result reads \cite{Moussallam:1994at,Dai:2019lmj}
\begin{align}\label{SDC}
\lim_{q^2\to\infty}\Gamma^{(a)}_{\mu\nu}(q,q)=2{\hat{d}}^{a\gamma\gamma}\frac{\langle {\bar{q}}q\rangle}{Q^4}(q_\mu q_\nu-q^2\eta_{\mu\nu})\,.
\end{align}
As opposed to what happened for the scalar transition form factors, the model displays the right asymptotic scaling. However, it is worth noting that, while ${\bar{{\cal{P}}}}^{(a)}$ and ${\bar{{\cal{Q}}}}^{(a)}$ separately match the momentum scaling predicted by the OPE, the ratio between the asymptotic coefficients, ${\bar{{\cal{P}}}}^{(a)}:Q^2{\bar{{\cal{Q}}}}^{(a)}$, which QCD sets at $3:1$, comes out approximately at $2:1$ and with a relative sign. This sign difference with the QCD prediction has the same origin as the one discussed above for $F_j^{(n,a)}$.  

We can also consider the limit where the vector momenta are hard and the scalar one is soft. In this case, to leading order, $q_1=-q_2\equiv q$ and one finds
\begin{align}
\lim_{q^2\to\infty}\Gamma^{(a)}_{\mu\nu}(q,-q)=\frac{64s_0\zeta}{15z_0^3}\frac{{\hat{d}}^{a\gamma\gamma}}{Q^4}(q_\mu q_\nu-q^2\eta_{\mu\nu})+{\cal{O}}(Q^{-6})\,.
\end{align}
Again, the scaling is the one expected from the OPE~\cite{Dai:2019lmj}.

Notice that Eqs. (\ref{SDCmodel}) and (\ref{SDC}) imply that the combination $s_0\zeta<0$. This condition is also found at low energies. By using 
\begin{align}
s(z,0)=z\left(1-\frac{\gamma}{\gamma+2}\frac{z^2}{z_0^2}\right)\,,
\end{align}
the first form factor yields 
\begin{align}
\lim_{q_j^2\to 0}{\cal{P}}^{(a)}(q_1,q_2)&=2\zeta s_0z_0{\hat{d}}^{a\gamma\gamma}\left(1-\frac{2}{3}\frac{\gamma}{\gamma+2}\right)\,.
\end{align}
The constant on the right-hand side is related to a NNLO coefficient in chiral perturbation theory, which has been estimated to be negative~\cite{Moussallam:1994at}. It is easy to prove that the term inside the parentheses above is positive for the phenomenologically acceptable values of $\gamma$, and therefore that $s_0\zeta<0$.
 
%%%%%%%%%%%%%%%%%%%%%%%%%%%%%%%%%%%%%%%%%%%%%%%%%%%%%%%%%%%%%%%%%
\section{Numerical analysis and discussion}
\label{sec:5}

In order to perform the numerical analysis, we need to define a strategy to fix the parameters of the model. From the starting action in Eq.~(\ref{action1}), one has nine parameters, namely the coefficients of the different bulk operators ($\lambda$, $c$, $\rho$, $\zeta_\pm$, $m_X$), the size of the fifth dimension $z_0$ and the parameters from the scalar boundary potential ($\mu$, $\eta$). However, for the scalar contributions to the HLbL, only a subset of them is relevant, namely $\rho$, $z_0$, the combination $\zeta=\zeta_++\tfrac{1}{2}\zeta_-$, $m_X$, and the parameters of the boundary potential, which can be traded for the quark condensate $\langle {\bar{q}}q\rangle$ and $\gamma$. As discussed in Sec.~\ref{sec:2}, in order to have conformal invariance asymptotically, it is convenient to fix the five-dimensional scalar mass $m_X$ to the value dictated by the AdS/CFT correspondence, $m_X^2=-3$.

We are therefore left with five independent parameters. The analysis of the previous section shows a possible strategy to fix them. We will require that $\zeta$ and $\rho$ match the $\langle SVV\rangle$ short-distance constraint of Eq.~(\ref{SDC}) and the decay width of the lowest-lying scalars into two photons, according to Eqs.~(\ref{widths}) and (\ref{F10}), where one should keep in mind that the constant ${\cal A}_n$ depends on $\rho$ (see Eq.(\ref{An})). 

Note, however, that our starting action is flavour symmetric. This means, in particular, that the lowest-lying scalar mesons come out degenerate in mass and they have the same decay width into two photons. Experimentally, the partial decay widths of the $f_0(980)$ and $a_0(980)$ are well known and given by \cite{Zyla:2020zbs}
\begin{align}\label{dataLE}
\Gamma_{\gamma\gamma}^{(f_0)}=0.29^{+0.11}_{-0.06}\,{\mathrm{keV}};\qquad  \Gamma_{\gamma\gamma}^{(a_0)}=0.30(10)\,{\mathrm{keV}}\,.
\end{align} 
For the $\sigma(500)$, however, rescattering effects are dominant and the determination is less precise. The most recent analyses suggest $\Gamma_{\gamma\gamma}^{(\sigma)}=(1.3-2)\,{\mathrm{keV}}$ \cite{Moussallam:2011zg,Dai:2014zta,Danilkin:2020pak}. 

\begin{figure}[t]
\begin{center}
\includegraphics[width=0.48\textwidth]{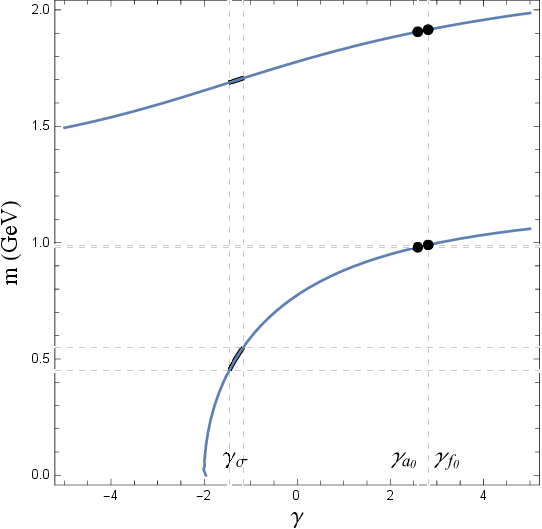}
\end{center}
\caption{Values of the parameter $\gamma_j$ for the different scalar masses. The intersections on the lower curve correspond to the $\sigma(500)$, $a_0(980)$ and $f_0(980)$ states. The intersections on the upper curve correspond to the first excited states. Similar curves exist for higher masses, defining the infinite towers of mesons of the model.} \label{fig:3}
\end{figure}

In order to be able to evaluate the scalar contributions to the HLbL in a realistic way, we need to introduce flavour breaking. We will adopt the strategy followed in Ref. \cite{Cappiello:2019hwh} for the Goldstone and axial-vector towers, and generate independent copies of the original Lagrangian for each of the different light scalar states. Accordingly, the parameters will in general have a flavour index. We will therefore have a threefold set of parameters associated to the three neutral scalar towers, whose lightest states are the $\sigma(500)$, $a_0(980)$ and $f_0(980)$.

The dynamical functions listed in (\ref{12Basis}) can then be expressed, using the spectral decomposition of the Green's function in Eq. (\ref{Greenspectral}), as
\begin{widetext}
\begin{align}\label{decomp}
&{\bar\Pi}_3(Q_1,Q_2,\tau)=\int_0^{z_0}dz\int_0^{z_0}dz'\left[{\cal{P}}^{(a)}_{12}(z)+(Q_1^2+Q_2^2+Q_1Q_2\tau){\cal{Q}}^{(a)}_{12}(z)\right]G_{(a)}(z,z';s){\cal{P}}^{(a)}_{34}(z')\nonumber\\
&=\sum_n \left[F_1^{(n,\sigma)}(Q_1,Q_2)+[Q_1^2+Q_2^2+Q_1Q_2\tau] F_2^{(n,\sigma)}(Q_1,Q_2)\right]\frac{1}{m_{n,\sigma}^2+Q_3^2}F_1^{(n,\sigma)}(Q_3,0)\nonumber\\
&+\sum_k \left[F_1^{(k,a_0)}(Q_1,Q_2)+[Q_1^2+Q_2^2+Q_1Q_2\tau] F_2^{(k,a_0)}(Q_1,Q_2)\right]\frac{1}{m_{k,a_0}^2+Q_3^2}F_1^{(k,a_0)}(Q_3,0)\nonumber\\
&+\sum_j \left[F_1^{(j,f_0)}(Q_1,Q_2)+[Q_1^2+Q_2^2+Q_1Q_2\tau] F_2^{(j,f_0)}(Q_1,Q_2)\right]\frac{1}{m_{j,f_0}^2+Q_3^2}F_1^{(j,f_0)}(Q_3,0)\,,
\end{align}
\end{widetext}
This result can be easily extended to the remaining functions in (\ref{12Basis}). 

For simplicity, we will assume a flavour-invariant quark condensate. The model is insensitive to the scale dependence of this quantity, and we will choose $\langle{\bar{q}}q\rangle=(-260\, {\mathrm{MeV}})^3$, which is consistent with the phenomenological determinations at 1 GeV. The size of the fifth dimension can be shown to be inversely proportional to the vector meson mass scale as \cite{Erlich:2005qh}
\begin{align}
z_0=\frac{\gamma_{0,1}}{m_{\rho}}\,,
\end{align}
where $\gamma_{0,1}$ is the first root of the Bessel function $J_0(x)$. We will fix $z_0=(322$ MeV$)^{-1}$, which guarantees that the first vector multiplet lies at $m_\rho=775$ MeV.

The remaining parameters, namely $\gamma$, $\rho$ and $\zeta$, will be flavour dependent. Using Eq. (\ref{spectrumS}), $\gamma$ can be fixed by the mass of the lightest scalars through
\begin{align}
\gamma_j=-m_j z_0 \frac{J_0(m_j z_0)}{J_1(m_j z_0)}\,,
\end{align}
where $m_j=m_\sigma, m_{f_0}, m_{a_0}$. In Fig. \ref{fig:3} we show the dependence of $\gamma$ with the scalar masses, where we have used $m_{a_0}=980(20)$ MeV and $m_{f_0}=990(20)$ MeV together with the conservative range $m_{\sigma}=(450-550)$ MeV \cite{Zyla:2020zbs}. The masses of the excited states are also predicted by this single parameter. However, the resulting spectrum for the excited states differs substantially from the physical one. 

The short-distance contraint of Eq. (\ref{SDC}) is a flavour-independent one, meaning that the ratio
\begin{align}\label{zetamatched}
\frac{\zeta_j}{\rho_j}&\simeq -\frac{1}{4}\left[\int_{\epsilon}^{\infty}dy y^6 K_1(2y)\left(K_1^2(y)-K_0^2(y)\right)\right]^{-1}\!\!\simeq -7.72
\end{align}
is flavour universal. Finally, combining Eqs. (\ref{widths}), (\ref{F10}) and (\ref{zetamatched}), one finds
\begin{align}\label{matchrho}
\frac{\rho_j}{({\hat{d}}^{j\gamma\gamma})^2}&=(7.72)^2\frac{8\pi\alpha^2}{m_j\Gamma_{j\gamma\gamma}}\frac{\langle{\bar{q}}q\rangle^2z_0^4\gamma_j^2}{\gamma_j^2+2\gamma_j+\omega_j^2}\nonumber\\
&\qquad\times\left[\frac{4J_3(\omega_j)-\omega_j J_4(\omega_j)}{J_0(\omega_j)}\right]^2\,,
\end{align}
where $\Gamma_{j\gamma\gamma}$ is the partial decay width of the lightest scalar in the tower. Note that it is the combination $\rho_j({\hat{d}}^{j\gamma\gamma})^{-2}$ which we determine. This means that the specific values of ${\hat{d}}^{j\gamma\gamma}$, i.e. the quark structure of the scalar mesons, is not specified. This is actually one of the advantages of our strategy for the determination of the free parameters. Strictly speaking, Eq. (\ref{matchrho}) only fixes the magnitude of $\rho_j$, but not its sign. From the positivity of the scalar two-point function and eq. (\ref{zetamatched}), one expects $\rho_j>0$ and $\zeta_j<0$. This is, however, irrelevant for the HLbL contribution, which is sensitive only to the square of $\rho_j$ and $\zeta_j$. 

The model has certainly limitations as a model of scalar mesons. The parameter $\rho$, for instance, has been determined in previous papers \cite{Erlich:2005qh,DaRold:2005vr,Cherman:2008eh} through the asymptotic behaviour of the two-point correlator
\begin{align}
\delta^{ab}\Pi^{SS}(q^2)&=i\int d^4x e^{iq\cdot x}\langle 0|T\left\{j^a(x)j^b(0)\right\}|0\rangle\,,
\end{align}
where $j^a(x)={\bar{q}}(x)t^a q(x)$. At large Euclidean momenta the OPE reads
\begin{align}
\lim_{Q^2\to \infty}\Pi^{SS}(Q^2)&=\frac{N_c}{16\pi^2}Q^2\log\frac{Q^2}{\mu^2}+{\cal{O}}(Q^{-2})\,.
\end{align}
Using the asymptotic expansions for the bulk-to-boundary propagators in Eqs.~(\ref{btobpropS}), our model successfully predicts the $Q^2 \log Q^2$ behaviour. However, the coefficient would be matched with $\rho=N_c(8\pi)^{-2}$, while Eq. (\ref{matchrho}) would predict $\rho_j\sim {\cal{O}}(10^{3})$. Such numerical mismatches at $N_c=3$ are, however, to be expected in a minimal model.\footnote{Similar tensions between low-energy phenomenology and short-distance constraints have already been pointed out in the literature, especially for five-dimensional fields that are dual to nonconserved four-dimensional currents. For a discussion, see Ref.~\cite{Domokos:2012da}.} We already mentioned above that the predicted spectrum of excited states matches poorly the physical spectrum. Additionally, the $g_{SPP}$ couplings of the lightest scalars with Goldstone pairs give results of ${\cal{O}}(10$ MeV$)$. This is the same range that was found in Ref. \cite{Colangelo:2008us} using a similar holographic model, and it falls short of the experimental results by 2 orders of magnitude. Despite these limitations, the strategy to fix the parameters is meant to ensure that the model is phenomenologically sound, at least for a determination of the scalar contribution to the HLbL. 

\begin{figure}[t]
\begin{center}
\includegraphics[width=0.45\textwidth]{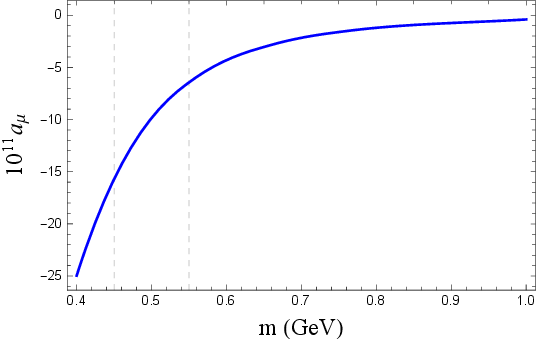}
\end{center}
\caption{$\sigma(500)$ contribution to the HLbL as a function of its mass for $\Gamma_{\sigma\gamma\gamma}=0.5$ keV.} \label{fig:4}
\end{figure}

\begin{figure}[t]
\begin{center}
\includegraphics[width=0.45\textwidth]{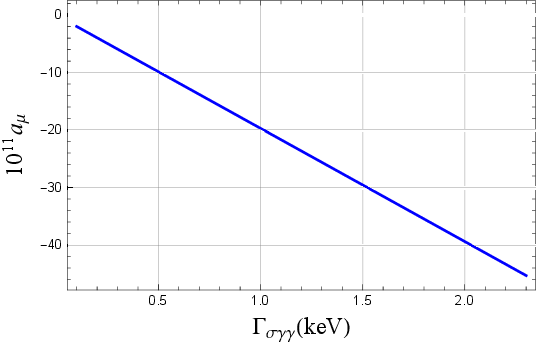}
\end{center}
\caption{$\sigma(500)$ contribution to the HLbL as a function of $\Gamma_{\sigma\gamma\gamma}$ for $m_\sigma=500$ GeV.} \label{fig:5}
\end{figure}

\begin{figure*}[t]
\begin{center}
\includegraphics[width=0.45\textwidth]{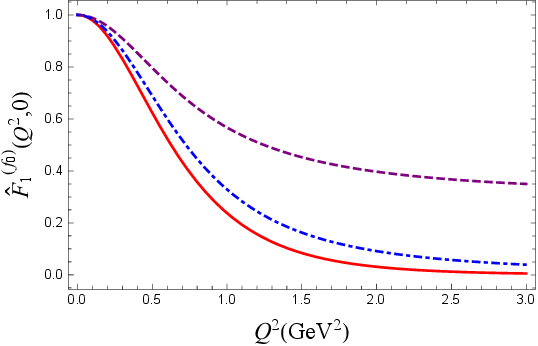}
\hspace{0.6cm}
\includegraphics[width=0.45\textwidth]{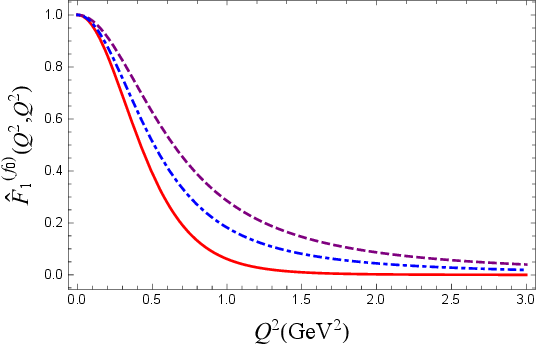}
\vskip 0.5cm
\includegraphics[width=0.45\textwidth]{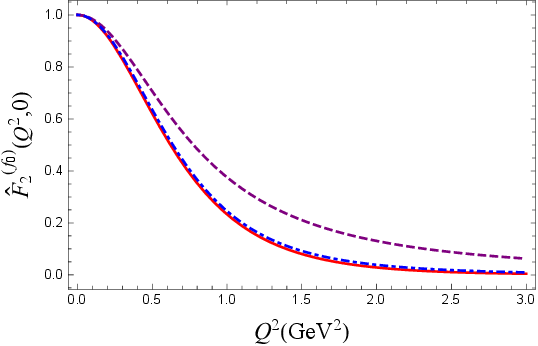}
\hspace{0.6cm}
\includegraphics[width=0.45\textwidth]{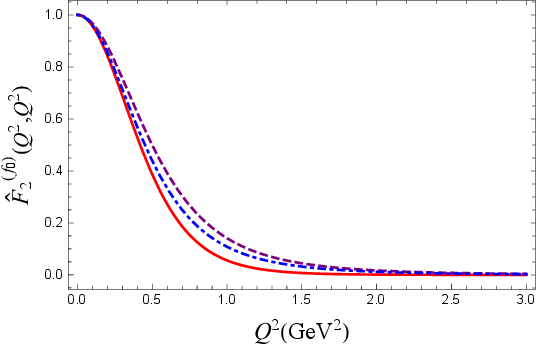}
\end{center}
\caption{Comparison of the normalized form factors ${\hat{F}}_j^{(n,a)}(q_1^2,q_2^2)\equiv F_j^{(n,a)}(q_1^2,q_2^2)[F_j^{(n,a)}(0,0)]^{-1}$ from our model (solid line), Ref.~\cite{Knecht:2018sci} (dashed line) and Ref.~\cite{Danilkin:2021icn} (dot-dashed line).} \label{fig:6}
\end{figure*}

The previous procedure to determine the parameters is well defined for the $a_0(980)$ and $f_0(980)$, which are rather narrow resonances. Instead, the $\sigma(500)$ cannot be considered a narrow state and one could argue that the description of this state in the model is not reliable. The most rigorous formalism to extract its parameters combines coupled Roy-Steiner equations with information from chiral perturbation theory (see, e.g., Refs. \cite{Moussallam:2011zg,Hoferichter:2011wk,Colangelo:2017fiz,Danilkin:2019opj,Danilkin:2020pak}). In our formalism, however, using the values for the mass and partial decay width extracted with this procedure would be inconsistent. To the best of our knowledge, the closest parametrization to our approach is the one of Ref. \cite{Filkov:1998rwz}, in which a Breit-Wigner model was employed to fit $\gamma\gamma\to \pi^0\pi^0$ data. The results of two different fitting strategies gave 
\begin{align}
m_\sigma&=547(45)\, {\mathrm{MeV}};\qquad \Gamma_{\sigma\gamma\gamma}=0.62(19)\, {\mathrm{keV}}\label{fit1}\,,\\
m_\sigma&=471(23)\, {\mathrm{MeV}};\qquad \Gamma_{\sigma\gamma\gamma}=0.33(07)\, {\mathrm{keV}}\label{fit2}\,,
\end{align}
where in the second line constraints from chiral perturbation theory were used. Of course, the interpretation of the partial widths above should be taken with care, but for our purposes this is not of relevance. The important point is that the parameters above fit the experimental data.

If one uses the strategy outlined above to determine the model parameters for the $\sigma(500)$ contribution, one finds $a_\mu^S=-8.11\cdot 10^{-11}$ and $a_\mu^S=-8.42\cdot 10^{-11}$, respectively, depending on whether one takes Eq. (\ref{fit1}) or Eq. (\ref{fit2}) as input. These values are not only compatible with each other but are on the ballpark of the one obtained from a dispersion relation analysis, $a_\mu^S=-9(1)\cdot 10^{-11}$ \cite{Colangelo:2017fiz,Danilkin:2021icn}. We take this as a strong indication that our strategy for the $\sigma(500)$ contribution is reliable.

It is difficult to come up with a reasonable estimate of the uncertainty associated to our number for the $\sigma(500)$ contribution to the HLbL. The mass and partial decay width in Eqs. (\ref{fit1}) and (\ref{fit2}) are strongly correlated. However, in Ref. \cite{Filkov:1998rwz}, no information is given of the correlation matrix. Our estimate
\begin{align}
a_\mu^S(\sigma)=(-8.5\pm 2.0)\cdot 10^{-11}
\end{align}    
is therefore orientative, but we believe that it correctly captures the right order of magnitude for the uncertainty.

In Figs. \ref{fig:4} and \ref{fig:5}, we have plotted the dependence of the HLbL on $m_\sigma$ (by keeping $\Gamma_{\sigma\gamma\gamma}=0.5$ keV) and $\Gamma_{\sigma\gamma\gamma}$ (by keeping $m_\sigma=500$ MeV), respectively. From Fig. \ref{fig:4}, it is clear that HLbL is most sensitive to $m_\sigma$ in the physical region ($450-500$ MeV), while the dependence fades away for higher values and is rather small for masses above $800$ MeV. We stress that the plot is done for fixed $\Gamma_{\sigma\gamma\gamma}$, which is not a realistic assumption. As a result, the spread of values for HLbL should not be taken in a quantitative sense. The plot is meant to illustrate a qualitative behaviour, but shows the importance of a precise determination of $m_\sigma$ and $\Gamma_{\sigma\gamma\gamma}$. 

Fig. \ref{fig:5} simply illustrates the linear dependence of the HLbL with $\Gamma_{\sigma\gamma\gamma}$, something that can be easily inferred from Eqs. (\ref{12Basis}), noticing that
\begin{align}
F_1^{(n,a)}(q_1^2,q_2^2)&=F_1^{(n,a)}(0,0)\frac{\omega_n^2}{z_0^5(4J_3(\omega_n)-\omega_n J_4(\omega_n))}\nonumber\\
&\times\int_\epsilon^{z_0}dz  z^4J_1(m_n z) v_1(z) v_2(z)\,.
\end{align}
Consequently, one finds that $F_1^{(n,a)}(q_1^2,q_2^2)\sim (\Gamma_{a\gamma\gamma}^{(n)})^{1/2}f(z_0,m_n,q_1^2,q_2^2)$, where $f$ is a function that can be computed numerically. A similar conclusion is reached for $F_2^{(n,a)}(q_1^2,q_2^2)$. 

Combining the two plots one can partially understand the much smaller result found for the $\sigma(500)$ contribution in Ref. \cite{Knecht:2018sci}, at least qualitatively. There, it was taken as input $m_\sigma=960(96)$ MeV and $\Gamma_{\sigma\gamma\gamma}=1.82(32)$ keV. Their result, taking into account only $F_1^{(n,a)}(q_1^2,q_2^2)$, was $a_\mu^S(\sigma)=-3.14\cdot 10^{-11}$. With those same input parameters, our model would yield $a_\mu^S(\sigma)=-1.89\cdot 10^{-11}$, which is roughly compatible. The main numerical discrepancy comes from the shape of the form factor, as we will discuss in the following.   

The contributions of $a_0(980)$ and $f_0(980)$ can be computed in a less problematic way. The main reasons for it are (i) both states are rather narrow, so no big deviations from the different parametrizations are expected (see e.g. the comments in Ref. \cite{Moussallam:2011zg}), and (ii) both states are heavy enough that the uncertainties on the mass are negligible. The contribution we find for both states is
\begin{align}
a_\mu^S(a_0)=-0.29(13)\cdot 10^{-11};\quad a_\mu^S(f_0)=-0.27(13)\cdot 10^{-11}\,,
\end{align}
which is in agreement with recent work~\cite{Knecht:2018sci,Danilkin:2021icn}, and compatible with earlier estimates~\cite{Pauk:2014rta,Danilkin:2016hnh} (see, however, the remarks in Ref. \cite{Danilkin:2021icn}). 

A comparison between numbers coming from Lagrangian and dispersive approaches is not straightforward, and nonpole terms can play a sizeable role, as emphasized in Ref. \cite{Danilkin:2021icn}. We have checked their impact by making the substitution $(Q_1^2+Q_2^2+Q_1Q_2\tau)\to \tfrac{1}{2}(Q_1^2+Q_2^2-m_n^2)$
in Eq. (\ref{decomp}) and the remaining scalar functions. This would yield $a_\mu^S(a_0)=-0.24(9)\cdot 10^{-11}$ and $a_\mu^S(f_0)=-0.22(11)\cdot 10^{-11}$, which is a rather modest shift. This can be understood by the numerical suppression of $F_2^{(1,a)}$ with respect to $F_1^{(1,a)}$ that our model exhibits.

In Fig. \ref{fig:6}, we compare the shape of the form factors for $f_0(980)$, normalized to their value at zero momenta, with those used in Refs. \cite{Knecht:2018sci,Danilkin:2021icn}. In the left and right panels, we plot $F_j^{(f_0)}(Q^2,0)$ and $F_j^{(f_0)}(Q^2,Q^2)$, respectively, in the region $Q^2\leq 3$ GeV$^2$. Discrepancies beyond that point are pinched by the HLbL kernels, which are rather peaked at very low $Q^2$. The relative impact of the form factors $F_j^{(n,a)}(q_1^2,q_2^2)$ was gauged in Ref. \cite{Knecht:2018sci} through the parameter $|\kappa_S|=m_{f_0}^2 F_2^{S}(0,0)[F_1^{(S)}(0,0)]^{-1}$. In our model, we find $|\kappa_{f_0}|=0.33$, while in \cite{Danilkin:2021icn} one would find $|\kappa_{f_0}|=0.65$. Ref. \cite{Knecht:2018sci} gave as benchmark points $\kappa_S=0$ and $\kappa_S=1$. In Fig. \ref{fig:6}, we have used the latter for illustration. The curves look very much alike, but our form factors are systematically smaller. This can be attributed to their short-distance behaviour, which is more suppressed than the one of QCD, as already discussed in the previous section. This short-distance behaviour is imprinted in the minimal action of Eq. (\ref{action1}). As discussed in Sec.~\ref{sec:4}, it could be improved with a more sophisticated action, something that is beyond the scope of the present paper.

In order to gauge the potential impact of short distances within the minimal action, we will modify the vacuum expectation value to $X_0={\cal{B}} z/z_0$, where ${\cal{B}}$ is a constant. According to the AdS/CFT correspondence, this behaviour is expected when quark masses are present. Here, we will take this as an {\it{ad hoc}} prescription, and fix the parameters in order to make sure that $\Gamma[f_0\to\gamma\gamma]$ and Eq. (\ref{SDC}) are fulfilled. With this prescription, the form factors scale asymptotically as $F_1(Q^2,Q^2)\sim Q^{-4}$ and $F_2(Q^2,Q^2)\sim Q^{-6}$, which is still unlike QCD but improved with respect to the model predictions. The results we find show a slight increase to $a_\mu^S(a_0)=-0.38(18)\cdot 10^{-11}$ and $a_\mu^S(f_0)=-0.36(18)\cdot 10^{-11}$. Interestingly, the function $f_1(w)$ would now agree with the one derived in Ref. \cite{Hoferichter:2020lap}, but not $f_2(w)$. Given the generic structure of the form factors, no matter the form of $X_0$, the model predicts $f_1(w)\neq f_2(w)$, in contrast to the results of Ref. \cite{Hoferichter:2020lap}.      

\begin{figure}[t]
\begin{center}
\includegraphics[width=0.48\textwidth]{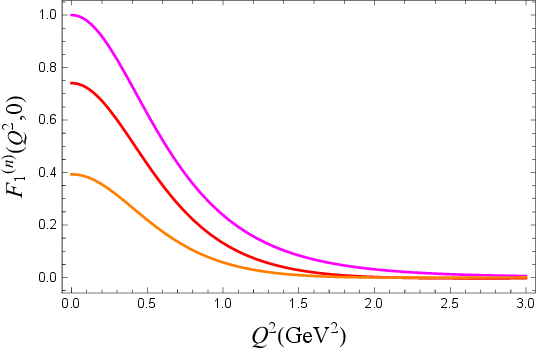}
\vskip 0.5cm
\includegraphics[width=0.48\textwidth]{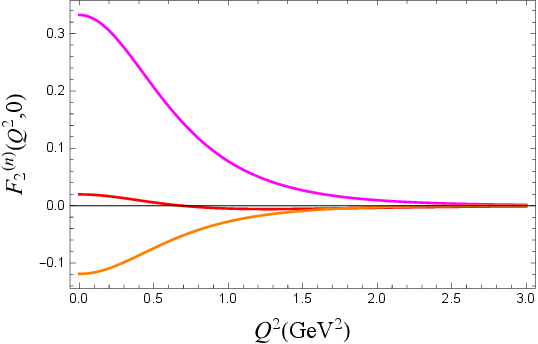}
\end{center}
\caption{Comparison of the form factors $F_1^{(n)}(q_1^2,q_2^2)$ (upper panel) and $F_2^{(n)}(q_1^2,q_2^2)$ (lower panel) for the $f_0(980)$ scalar resonance (magenta curves) and the first two excitation states (red and orange curves, respectively). All form factors are normalized to $F_1^{(f_0)}(0,0)$.} \label{fig:7}
\end{figure}

The model also provides a prediction for the contribution of the whole tower of scalar excitations. The results for the first excitation states together with the total contributions of the three scalar towers are shown in Table~\ref{tab:1}. They amount to an overall $3\%$ correction to the sub-GeV scalar contribution. Interestingly, when one considers $X_0={\cal{B}} z/z_0$, this number is actually 1 order of magnitude smaller, despite the fact that the asymptotic behaviour is enhanced. This can be traced back to the interplay of the form factors. We have already mentioned in the previous section that one of the limitations of the present model is that asymptotically $F_2^{(n,a)}(q_1^2,q_2^2)$ comes out with the same sign as $F_1^{(n,a)}(q_1^2,q_2^2)$, as opposed to what QCD predicts. However, the relative sign and weight at intermediate and low energies depends on the excitation mode. In Fig. \ref{fig:7}, we illustrate this feature for the $f_0(980)$ and the first two excitation states. With the choice $X_0={\cal{B}} z/z_0$, the contribution of the first excited state turns out to be extremely suppressed. The remaining states have a negligible contribution.  

The numerical impact of the asymptotic region can also be evaluated by comparing the contribution of the whole towers with the sum over discrete states. The scalar functions defined in Eqs.~(\ref{scalarF}) 
scale asymptotically as ${\hat{\Pi}}_4\sim Q^{-14}$ and ${\hat{\Pi}}_{17}\sim Q^{-16}$ for each single-particle scalar contribution. When the whole towers are taken into account, the behaviour gets enhanced to ${\hat{\Pi}}_4\sim Q^{-10}$ and ${\hat{\Pi}}_{17}\sim Q^{-12}$. These scalings are still suppressed with respect to the single-particle expressions used in Refs. \cite{Knecht:2018sci,Danilkin:2021icn} and to the OPE results reported in Refs. \cite{Bijnens:2019ghy,Bijnens:2020xnl,Bijnens:2021jqo}. However, they can be a guidance to understand how much enhancement one could expect from changing the large-$Q^2$ behaviour. In Table~\ref{tab:1}, ones sees that the contribution of the tower is mostly saturated by the first two states. We therefore conclude, as could be expected from the peaked shape of the kinematical kernels, that the behaviour of the dynamical functions beyond a few GeV is of little numerical importance.  

\begin{table}[t]
\begin{center}
\begin{ruledtabular}
\begin{tabular}{cccc}
 & $n=1$ & $n=2$ & $\mbox{\bf Total}$  \\
\hline
\\
$a_\mu^{\rm{S}}(\sigma)$ & -8.5(2.0) & -0.07(2) & -8.7(2.0) \\
\\
$a_\mu^{\rm{S}}(a_0)$ & -0.29(13) & -0.025(10) & -0.32(14) \\
\\
$a_\mu^{\rm{S}}(f_0)$ & -0.27(13) & -0.025(9) & -0.29(14) \\
\\
\hline
\\
$a_\mu^{\rm{S}}$ & -9(2) & -0.12(4) & -9(2) \\
\end{tabular}
\end{ruledtabular}
\caption{\label{tab:1} Results for the scalar contributions to $a_{\mu}^{\rm{HLbL}}\;\times\;10^{11}$ for the set of values described in the main text.}
\end{center}
\end{table} 

Taking all the previous points into account, we estimate the scalar contribution to the HLbL at
\begin{align}
a_\mu^S=-9(2)\cdot 10^{-11}\,,
\end{align}
with roughly $90\%$ of this number coming from the $\sigma(500)$. The error budget is likewise dominated by the $\sigma(500)$, whereas scalar states above the GeV scale contribute no more than $15\%$ to the total uncertainty. This number is compatible with the one estimated with the Nambu-Jona-Lasinio model, $a_\mu^S=-7(2)\cdot 10^{-11}$ \cite{Bijnens:1995cc,Bijnens:1995xf}, with dispersive analyses \cite{Danilkin:2021icn}, and also in agreement with the values reported in Refs. \cite{Jegerlehner:2009ry,Prades:2009tw,Jegerlehner:2017gek}.

%%%%%%%%%%%%%%%%%%%%%%%%%%%%%%%%%%%%%%%%%%%%%%%%%%%%%%%%%%%%%%%%%
\section{Conclusions}
\label{sec:6}

In this paper, we have provided an estimate of the scalar contribution to the HLbL, including the $\sigma(500)$, $a_0(980)$ and $f_0(980)$ states together with an infinite tower of excited scalar states with a toy model. This toy model of hadronic QCD is most easily built with a five-dimensional setting using holographic techniques.   

Our final result reads $a_\mu^S=-9(2)\cdot 10^{-11}$, where a conservative estimate for the uncertainty is given. This includes the uncertainty on the $\sigma(500)$ parameters, which overwhelmingly dominate, together with the effects of the other light sub-GeV states together as well as the infinite towers of excited states. Our result agrees with previous inclusive scalar estimates and points at a neatly negative contribution for the scalar contribution to the HLbL.  

One of the advantages of the model is that it is minimal and accordingly has a small number of free parameters. However, this also entails some limitations. The most prominent of these affect the scalar transition form factors, with some mismatches with QCD expectations. We have argued that these shortcomings have a limited impact on the HLbL and are in any case taken into account in the final error band. 

The estimate of the contribution of scalar resonances beyond 1 GeV is in general hindered by the rather uncertain knowledge of their couplings to two photons. For the model, both the poles and couplings of the whole spectrum are determined only by two parameters, which are generated by the lowest-lying resonance on each flavour channel. This leads to a rather poor description of the states populating the 1-2 GeV energy window, and one cannot exclude that these states contribute in a more significant way than the one estimated in this paper. Additionally, the spectrum of the excited scalar states in the model is more sparse than the QCD one, which also points at a possible enhanced contribution. These limitations of the model are also considered in the final error band. 

It is tempting to add the value reported here for the scalar contributions to the one obtained in Ref. \cite{Cappiello:2019hwh} for the Goldstone and axial-vector contributions using a similar approach. If these numbers are added naively, and the errors are added linearly, one would find $a_\mu^{\rm{HLbL}}=116(17)\cdot 10^{-11}$, which is in agreement with all the recent estimates of the HLbL. However, this number has to be taken with care, at least for two reasons. First, the previous estimate does not contain the pion and kaon loop contributions, together with other 1/$N_c$-suppressed contributions. With the Goldstone loops taken from Refs. \cite{Colangelo:2017fiz,Colangelo:2017qdm,Eichmann:2019bqf}, our estimate for the HLbL would instead hover around $a_\mu^{\rm{HLbL}}=100(17)\cdot 10^{-11}$. The second point to take into consideration is that the five-dimensional model used in Ref. \cite{Cappiello:2019hwh} contained only the gauge sector. The addition of scalar fields into the action, as we did in this paper, not only affords an estimate of the scalar-exchange contributions, but it also has an effect on the axial-vector and Goldstone sectors. In particular, the presence of a scalar sector makes it possible to have realistic values for both the Goldstone decay constants and $m_{\rho}$, something that could not be achieved in Ref. \cite{Cappiello:2019hwh}. This will affect the estimate of the axial-vector and Goldstone contributions to the HLbL reported in Ref. \cite{Cappiello:2019hwh} and it calls for a reanalysis of Goldstone, axial-vector and scalar exchange using the action of Eq.~(\ref{action1}). This is left for a future publication. 

%%%%%%%%%%%%%%%%%%%%%%%%%%%%%%%%%%%%%%%%%%%%%%%%%%%%%%%%%%%%%%%%%
\section*{Acknowledgments}

We thank Martin Hoferichter, Marc Knecht, Josef Leutgeb, Anton Rebhan and Peter Stoffer for correspondence. 
L.C. and G.D. were supported in part by the INFN research initiative Exploring New Physics (ENP). The work of O.C. is supported in part by the Deutsche Forschungsgemeinschaft (DFG, German Research Foundation) under Grant No.
396021762 - TRR 257 "Particle Physics Phenomenology after the Higgs Discovery."

%%%%%%%%%%%%%%%%%%%%%%%%%%%%%%%%%%%%%%%%%%%%%%%%%%%%%%%%%%%%%%%%%

\end{document}